 \documentclass[preprint,showpacs,preprintnumbers,superscriptaddress,amsmath,amssymb]{revtex4}
 \usepackage{cases}

 \usepackage{mathrsfs}
 \usepackage{amssymb}
 \usepackage[dvips]{graphicx}
 \usepackage{dcolumn}
 \usepackage{bm}
 \usepackage{subfigure}
 \usepackage{floatflt}
 \usepackage{float}
 \usepackage{rotating}
 \usepackage{multirow}
 \usepackage[bookmarksnumbered,bookmarksopen,colorlinks,citecolor=blue,linkcolor=blue]{hyperref} %
 \usepackage{booktabs}
 \usepackage{epsfig,graphics,psfrag,amsmath,amssymb}
 \usepackage{sidecap}

\begin{document}
    \preprint{00-000}

    \title{Further study on mechanism of production of light complex particles in nucleon-induced reactions}

    \author{Dexian Wei}%
    \affiliation{College of Physics and Technology, Guangxi Normal University, Guilin, 541004, People's Republic of China}
    \author{Lihua Mao}%
    \affiliation{College of Physics and Technology, Guangxi Normal University, Guilin, 541004, People's Republic of China}
    \author{Ning Wang}%
    \affiliation{College of Physics and Technology, Guangxi Normal University, Guilin, 541004, People's Republic of China}
    \author{Min Liu}%
    \affiliation{College of Physics and Technology, Guangxi Normal University, Guilin, 541004, People's Republic of China}
    \author{Li Ou}
    \email{only.ouli@gmail.com}
    \affiliation{College of Physics and Technology, Guangxi Normal University, Guilin, 541004, People's Republic of China}

    \date{\today}

    \begin{abstract}

    The Improved Quantum Molecular Dynamics (ImQMD) model incorporated with the statistical decay model
    is used to investigate the intermediate energy nucleon-induced reactions.
    In our last work, the description on light complex particle emission has been great improved with
    a phenomenological mechanism called surface coalescence and emission introduced into ImQMD model.
    In this work, taking account of different specific binding energies and separation energies
    for various light complex particles, the phase space parameters in surface coalescence model are readjusted.
    By using the new phase space parameters set with better physical fundament,
    the double differential cross sections of emitted light complex particles are found to be in
    better agreement with experimental data.
    \end{abstract}

    \pacs{25.40.Sc, 24.10.-i, 25.70.Gh}
    \keywords{spallation reactions, light complex particles' production, transport
                model and statistical decay model analysis}
    \maketitle



\section{Introduction}

    Since spallation reactions were investigated for the first time 80 years ago by using cosmic rays\cite{Rossi},
    spallation reactions have gained interests in fundamental and applied research fields.
    As optimum neutron source\cite{Bauer}, or for energy production and nuclear waste transmutation
    in accelerator-driven system\cite{Bowma,Abder}, it recalls the attentions in study on spallation reactions.
    These applications require a large amount of spallation reaction data\cite{Titar},
    which can not be provided all by experiments\cite{NEA98}.
    So a theoretical model with powerful prediction ability is imperative.
    The International Atomic Energy Agency (IAEA) and the Abdus Salam International Center for Theoretical Physics
    have recently organized twice international workshop to make comparisons of spallation models and codes,
    including PHITS\cite{Iwase}, BUU\cite{Berts}, QMD\cite{Aiche}, JAM\cite{Nara1},
    JQMD\cite{Niit1}, INCL4\cite{Boud1}, ISABEL\cite{Yariv}, Bertini\cite{Bert1,Bert2}, Geant\cite{Agost},
    IQMD\cite{Hartnack}, RQMD\cite{Lehmann}, TQMD\cite{Puri} et al,
    by merged with various statistical decay models such as GEM\cite{Furihata00,Furihata02},
    GEMINI\cite{Chari}, ABLA\cite{Junghans} et al. The productions of the neutron, proton, pions and isotopes
    can be overall described well by the most of given models\cite{Titar}.
    But the data for the light complex particles (LCPs), i.e. $^2$H, $^3$H, $^3$He, and $^4$He,
    can not be reproduced well.
    Only models which have a specific mechanism to emit energetic clusters,
    such as coalescence during the intranuclear cascade stage or pre-equilibrium emission of composite nuclei,
    can reproduce the high-energy tail of LCPs double differential cross sections (DDXs)\cite{Lera1}.
    The information of yield of hydrogen and helium element is quite important for design of nuclear project
    including target and shield.
    Therefore, the description of model on LCPs emission should be improved to satisfy the request of
    the applications\cite{Lera1,DAVID}.

    In our previous works, by applying the Improved Quantum Molecular Dynamics (ImQMD) Model merged statistical decay model,
    a series of studies on the proton-induced spallation reactions
    at intermediate energies have been made\cite{Liou1, Liou2, Liou3}.
    Nuclear data including neutron DDXs, proton DDXs, mass, charge and isotope distributions can be overall reproduced quite well.
    In our last work\cite{Dexian}, with a phenomenological mechanism called surface coalescence and emission
    is introduced into ImQMD model,
    the description on LCPs produced in nucleon-induced reactions are great improved but still not good enough.
    The motivation of this work is to further improve the surface coalescence and emission model.
    We expect the model with new parameters can describe the experimental data more accurately.

    The paper is organized as follows. In the next section,
    we make a brief introduction on the model.
    In section III, we give the calculation results and make some discussion.
    Finally, a summary is given in Sec. IV.

    \section{Mechanism of surface coalescence}

    ImQMD model merged GEM2\cite{Furihata00,Furihata02}, the same method as one used in our last work\cite{Dexian},
    is used in this work.
    The very detailed introduction on the ImQMD model can be found in ref. \cite{Dexian} and the reference therein.
    Here we only make an introduction on the surface coalescence mechanism which has been introduced into ImQMD05 model.
    After the incident nucleon has touched the target nuclei to form the compound nuclei,
    we define a sphere core with radius $R_0$ surrounding by a surface with width $D_0$.
    At each time step, any fast nucleon passing the surface region to leave
    the compound system is taken as leading nucleon.
    An inspection is made over all other regular nucleons in order to check whether
    there are one or several nucleons close enough in phase space to allow
    the formation of a stable composite-particle satisfying the following condition
    \begin{equation}\label{condition}
    R_{im}\times P_{im}\leq h_0 \hspace{0.5cm}
    {\rm{with}} \hspace{0.5cm} R_{im}\geq 1~\rm{fm},
    \end{equation}
    Where the $R_{im}$ and $P_{im}$ are the Jacobian coordinates of the $i$th nucleon,
    i.e., the relative spatial and momentum coordinates of the considered $i$th nucleon with respect to the subgroup cluster $m$.
    The value of $h_0$ is adjustable by fitting the experimental data.
    And the condition $R_{im}\geq 1~\rm{fm}$ gets rid of the unreal LCPs
    constituted by the nucleons being too close to each other,
    due to the repulsion of nucleon-nucleon interaction in short distance\cite{Letou}.
    In present work, the following LCPs are considered: $^2$H, $^3$H, $^3$He, and $^4$He.
    By the method of LCPs to be constructed, candidate nucleon belongs to a heavier LCP
    also belongs to a lighter LCP. So LCPs are checked to be emitted according to the priority list:
    $^4$He$>^3$He$>^3$H$>^2$H, say the heavier LCP is first tested for emission,
    as the same order as that in Ref.\cite{Letou,Boud2,Watan}.
    Finally the candidate LCP can be emitted or not depends on whether its kinetic energy
    is high enough to tunnel through the Coulomb barrier. The kinetic energy of the LCPs can be calculated as
    \begin{equation}
    E_{\rm{lcp}}=\sum_{i=1}^{A_{\rm{lcp}}} (E_i+V_i)+B_{\rm{lcp}}
    \end{equation}
    Where the $E_i$ and $V_i$ are the kinetic energy and the potential energy of
    the $i$th constituent nucleon, respectively, $A_{\rm{lcp}}$ and $B_{\rm{lcp}}$ are the mass number
    and the binding energy of the LCP, respectively. If all conditions are meet,
    the LCP is emitted in the direction of its c.m. momentum.
    Otherwise, all nucleons in the ``LCP'' are set free and become available again in the nucleus and in the ImQMD process,
    the leading nucleon is emitted as a free nucleon.
    In this coalescence model, composite-particles are thus not allowed being formed in
    the interior of the nucleus but only in the surface layer.
    It is reasonable according to the knowledge from nuclear structure and reaction.
    For each leading nucleon, the LCP formation and emission are tested in the priority list.
    At each time step, the same test is repeated until the end of ImQMD simulation.

\section{Results and discussion}

\begin{figure*}[h]
    \includegraphics[height=5cm]{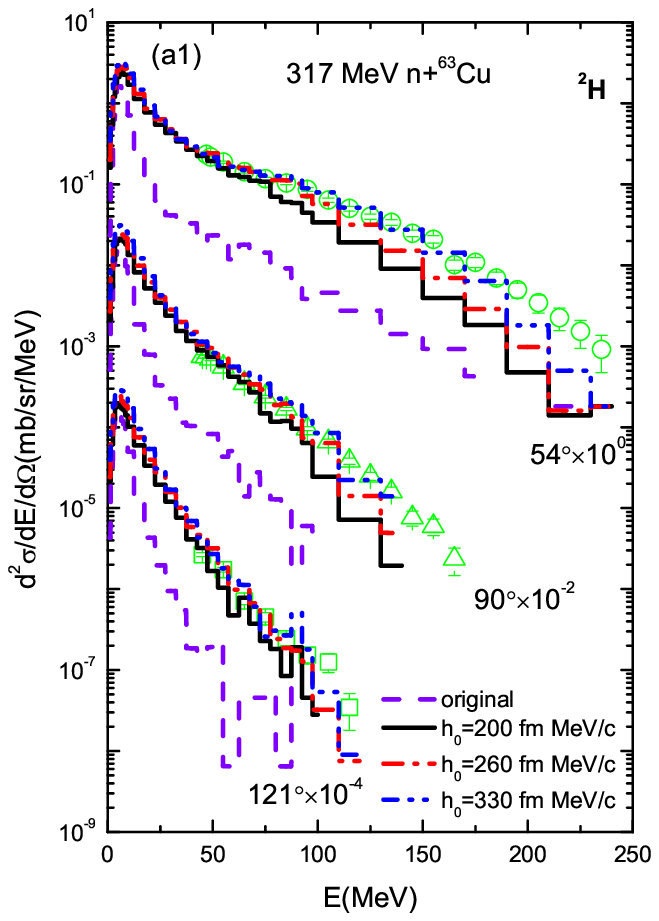}
    \includegraphics[height=5cm]{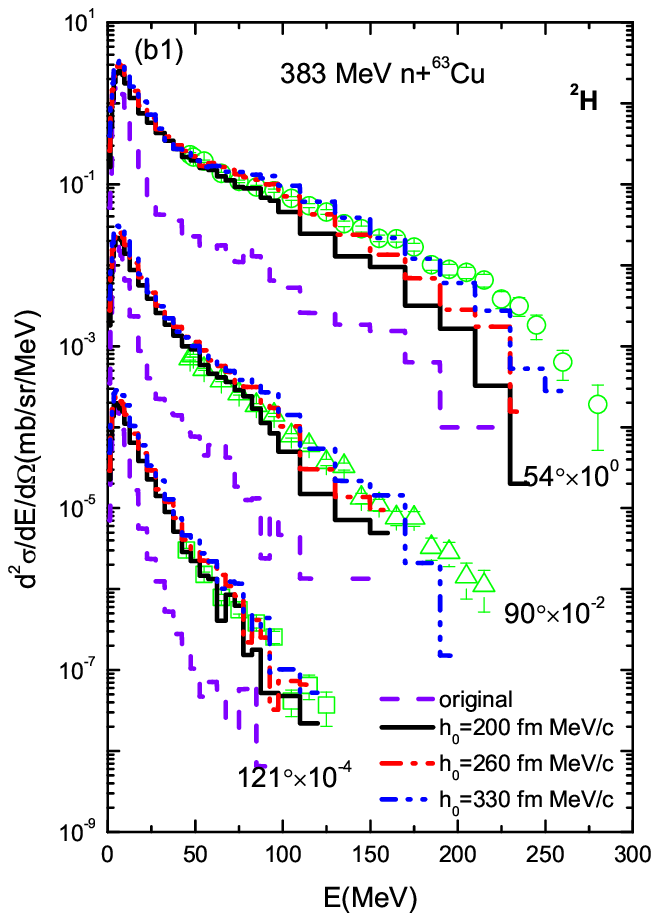}
    \includegraphics[height=5cm]{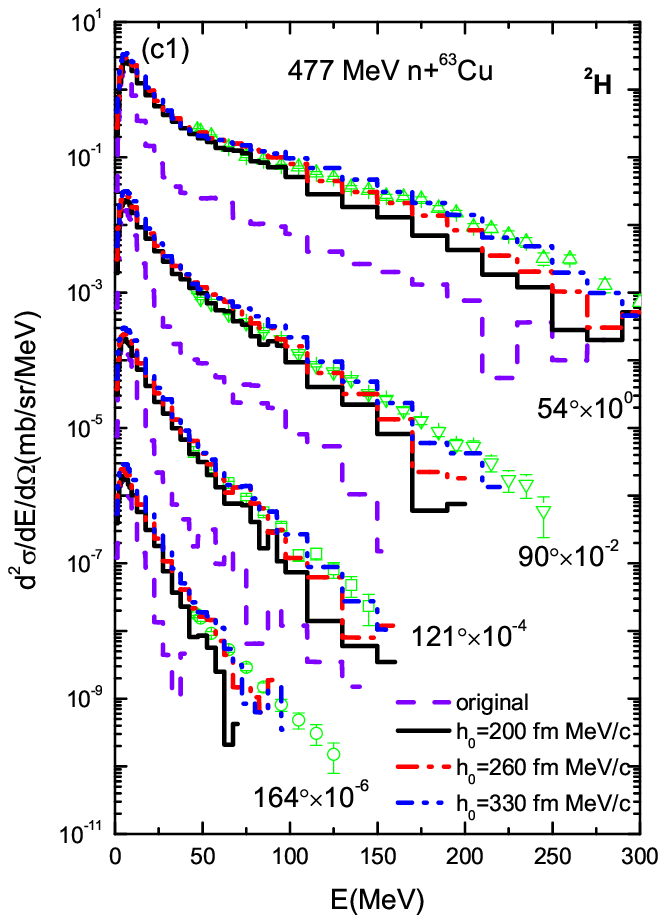}
    \includegraphics[height=5cm]{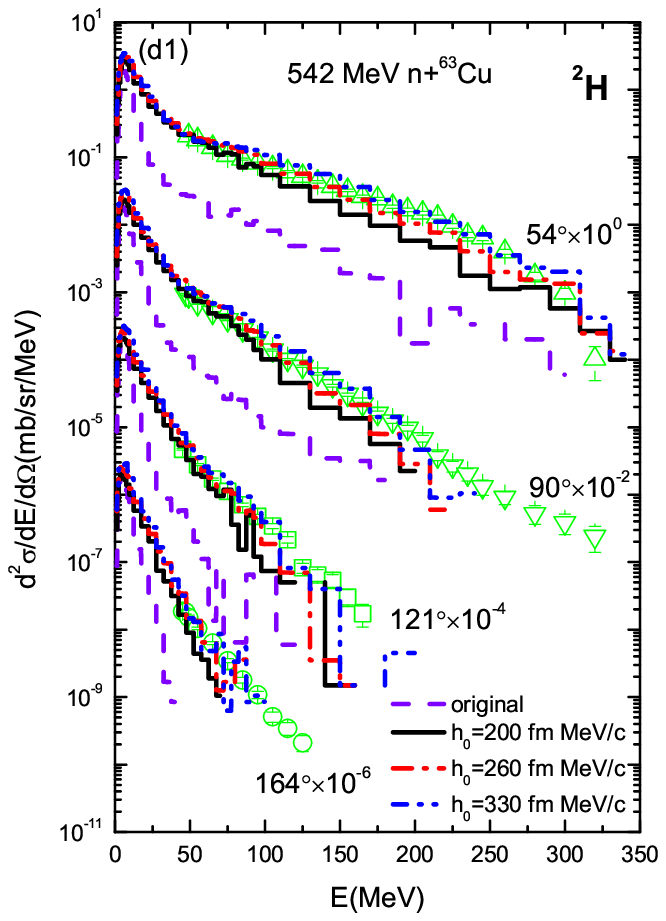}\\
    \includegraphics[height=5cm]{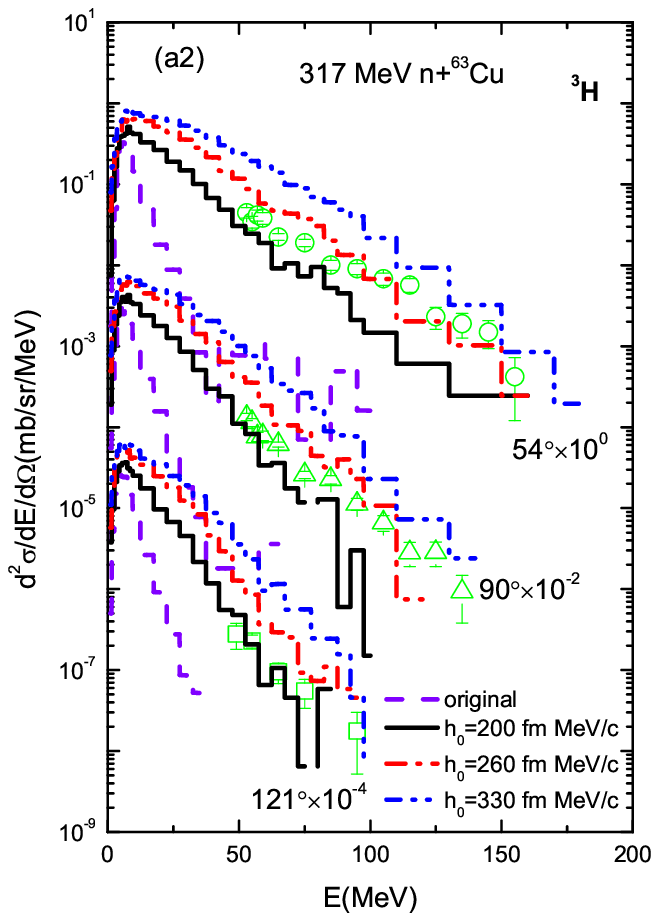}
    \includegraphics[height=5cm]{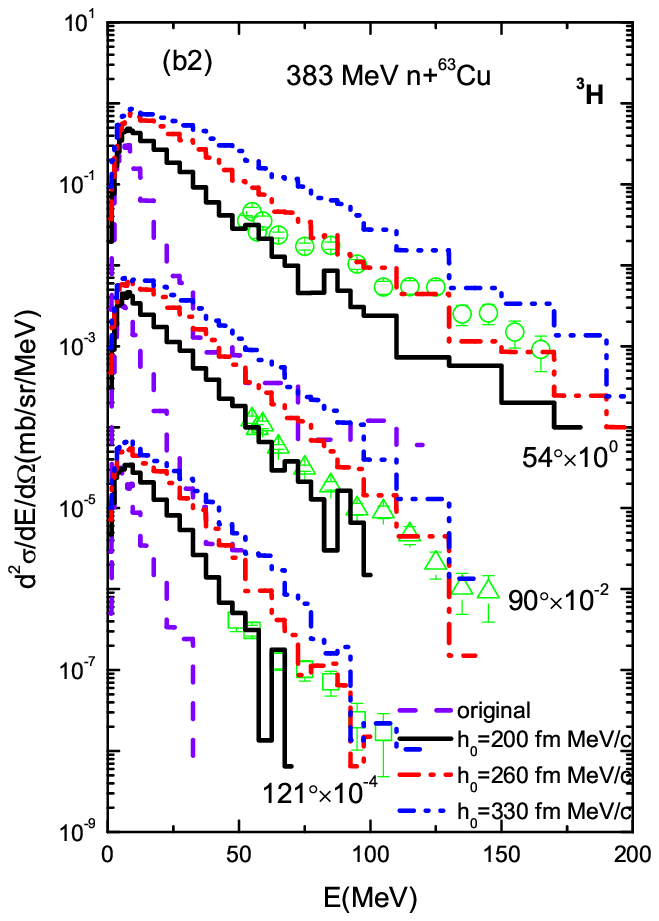}
    \includegraphics[height=5cm]{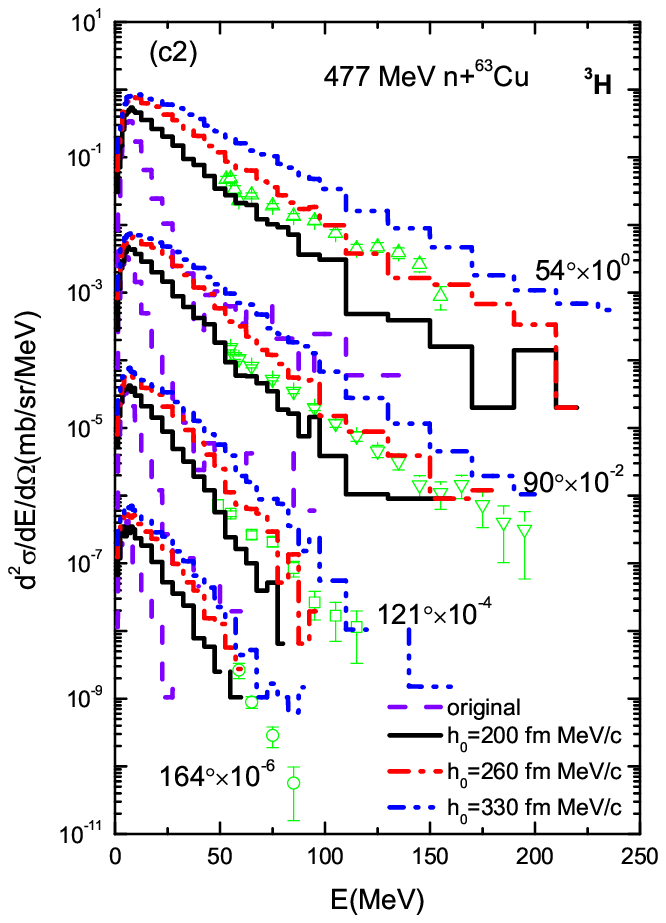}
    \includegraphics[height=5cm]{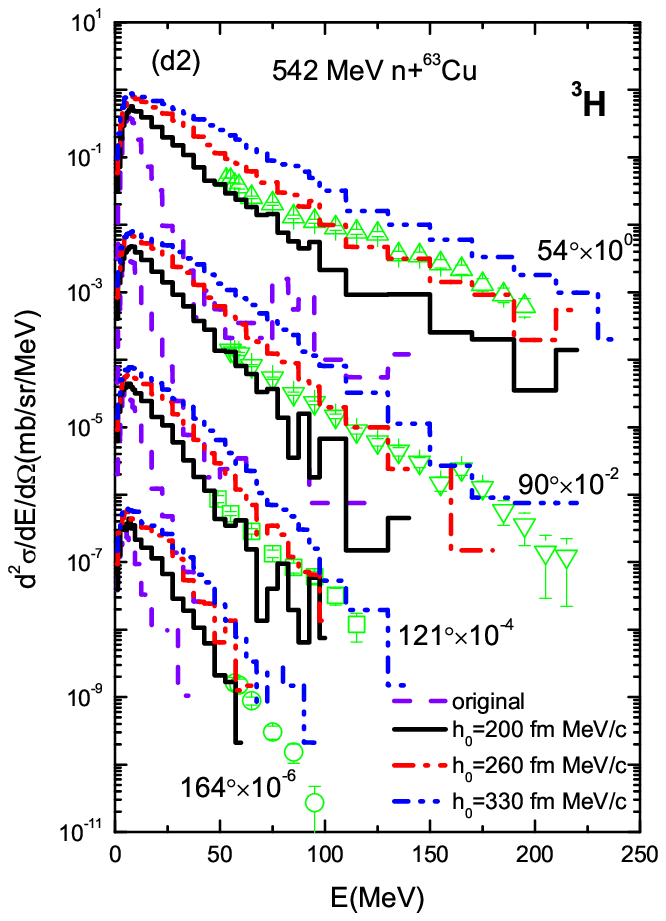}
    \caption{(Color online) Calculated DDXs of $^2$H and $^3$H in the reactions
    $n+^{63}$Cu at 317, 383, 477 and 542 MeV. $h_0$=200, 260 and 330 fm MeV/$c$
    are adopted, respectively. Experimental data are taken from \cite{Franz}.}
    \label{fig1}
    \end{figure*}

    By systematic comparison between calculation results and
    experimental data of nucleon-induced reactions,
    the parameters in the surface coalescence model are fixed.
    One best choice is $R_0$= 1.4A$^{1/3}$, $D_0$=2.3 fm, and
    phase space parameters
    \begin{numcases}
    {h_{0}=}
    200~{\rm fm~MeV}/c, &$E_{\rm{lab}}\leq 300$ MeV,\nonumber\\
    260~{\rm fm~MeV}/c, &$300~{\rm{MeV}} < E_{\rm{lab}}\leq 500$ MeV, \label{setI}\\
    330~{\rm fm~MeV}/c, &$E_{\rm{lab}}> 500$ MeV.\nonumber
    \end{numcases}
    We call this phase space parameters set as setI.
    Then with the fixed parameters,
    chosen once for all, the prediction power of the model is tested
    by the nucleon-induced reactions on various targets with energies from 62 to 1200 MeV.
    And it is found that, with surface coalescence mechanism introduced into ImQMD model,
    the description on the DDXs of LCPs is great improved.
    The experimental data can be overall reproduced well.
    But the calculation results are still a little far from satisfactory in detail.
    The experimental data of each single LCP emitted in the same reaction
    can not be simultaneously and satisfactorily reproduced with one given $h_0$.
    For example, calculated DDXs of $^2$H and $^3$H in the reactions
    $n+^{63}$Cu at 317, 383, 477 and 542 MeV are represented in Fig. \ref{fig1},
    with $h_0$=200, 260 and 330 fm MeV/$c$ adopted, respectively.
    Calculation results without surface coalescence mechanism are also shown in the figure.
    As a rule in this article and for the sake of clarity,
    DDXs are displayed after multiplication
    by 10$^0$, 10$^{-2}$, 10$^{-4}$, 10$^{-6}$, etc., as noted in the figures.
    One can see that, for $^2$H, larger $h_0$ is required to reproduce the experimental data.
    But for $^3$H, calculations with large $h_0$ overestimates the experimental data.
    Finally, the compromised values of $h_0$ listed in Eq. \eqref{setI} are selected to make calculations
    to overall reproduce experimental data.
    So it seems that, to reproduce the experimental data of each LCPs as well as possible,
    various $h_0$ should be adopted when construct different LCPs.

    From the knowledge of nuclear structure, we know that the nuclei
    with higher specific binding energy is more stable.
    In table \ref{Table1}, the specific binding energies for LCPs are represented.
    The descending order of specific binding energies for LCPs is $^4$He$>^3$H$>^3$He$>^2$H.
    It means that the nucleons in $^4$He are bound most tightly and the nucleons in $^2$H are bound most loosely.
    On the other hand, separation energy also reflects the tightness between the nucleons in LCP.
    From Table \ref{Table1}, one can see that the descending order of separation energies
    for LCPs is $^4$He$>^3$H$>^3$He$>^2$H.
    It means that, compared to nucleon in $^2$H, the last nucleon in $^4$He is more difficult to
    be separated from mother nuclei.
    In surface coalescence model, one by one candidate nucleon is coalesced to leading nucleon to form LCP,
    so the condition Eq. \eqref{condition} for the last candidate nucleon should be adjusted according to
    the effect of specific binding energies and separation energies of LCPs.
    The LCP with large specific binding energy and separation energy requires
    the last candidate nucleon to be more close to the subgroup cluster in the phase space with smaller $h_0$.
    With $R_0$= 1.4A$^{1/3}$, $D_0$=2.3 fm, as same as ones used in last work, the value of $h_0$ are determined
    by fitting the DDXs experimental data of $^2$H, $^3$H, $^3$He, $^4$He
    in the nucleon-induced reactions on various targets with incident energies from 62 MeV to 1.2 GeV.
    In table \ref{Table1}, we list the value of $h_0$ used to construct different LCPs.
    For example, a leading proton can pick any neutron satisfying $R_{im}\times P_{im}\leq 300$ fm MeV/$c$ to form $^2$H.
    But for a generated $^3$He to form $^4$He, it must pick up the last neutron satisfying $R_{im}\times P_{im}\leq 200$ fm MeV/$c$.
    According to $S_n(^3{\rm{H}})>S_p(^3{\rm{He}})$, $h_0(^3{\rm{H}})$ should be smaller than $h_0(^3{\rm{He}})$.
    But one can see that $h_0(^3{\rm{H}})>h_0(^3{\rm{He}})$ is taken,
    because there is only difference of 0.76 MeV in separation energies between two particles,
    and the emission priority $>^3$He$>^3$H is also a factor to be considered.
    The similar phase space parameter set are used in Ref. \cite{Letou}.
    We call this phase space parameters set as setII.
    \begin{table*}
    \tabcolsep 0pt
    \caption{\label{Table1}
    Specific binding energies, separation energies and value of $h_0$ for LCPs, respectively.}
    \vspace*{-24pt}
    \begin{center}
    \def\temptablewidth{0.7\textwidth}
    {\rule{\temptablewidth}{1pt}}
    \begin{tabular*}{\temptablewidth}{@{\extracolsep{\fill}}ccccc}
        LCP     & construction     & $E_{\rm{B}}/A$ (MeV)   & $S_{n,p}$ (MeV)   &$h_0$(fm MeV/$c$)\\ \hline
        $^2$H   & $p+n$            & 1.1                    & 2.24              &300 \\
        $^3$H   & $n+^2$H          & 2.9                    & 6.24              &260 \\
        $^3$He  & $p+^2$H          & 2.6                    & 5.48              &230 \\
        $^4$He  & $p+^3$H          & 7.1                    & 19.82             &200 \\
        $^4$He  & $n+^3$He         & 7.1                    & 20.58             &200
    \end{tabular*}
    {\rule{\temptablewidth}{1pt}}
    \end{center}
    \end{table*}

\begin{figure*}[htp]
    \begin{center}
    \includegraphics[width=1\textwidth]{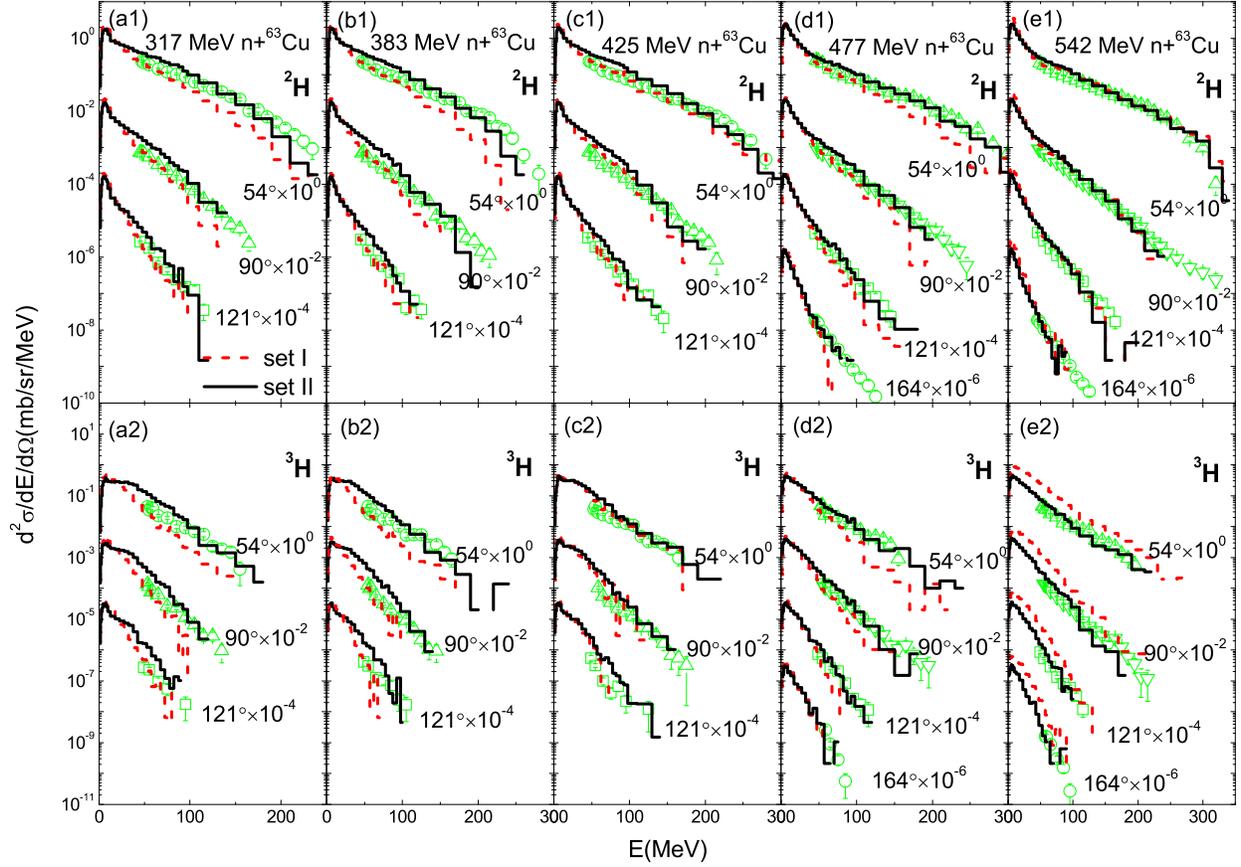}
    \caption{(Color online) Calculated DDXs of $^2$H and $^3$H in the reactions
    $n+^{63}$Cu at 317, 383, 425, 477 and 542 MeV with parameter setI and setII adopted,
    respectively. Experimental data are taken from \cite{Franz}.}
    \label{fig2}
    \end{center}
    \end{figure*}

\begin{figure*}[htp]
    \begin{center}
    \subfigure{\label{fig3a}
    \includegraphics[height=4.5cm]{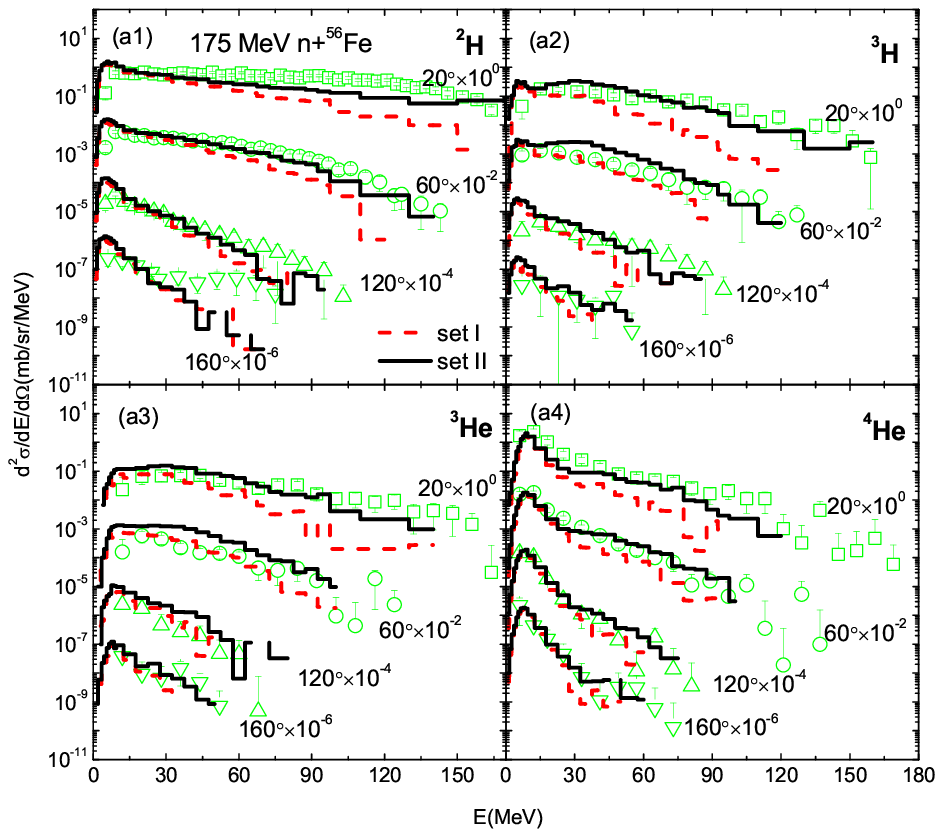}}
    \subfigure{\label{fig3b}
    \includegraphics[height=4.5cm]{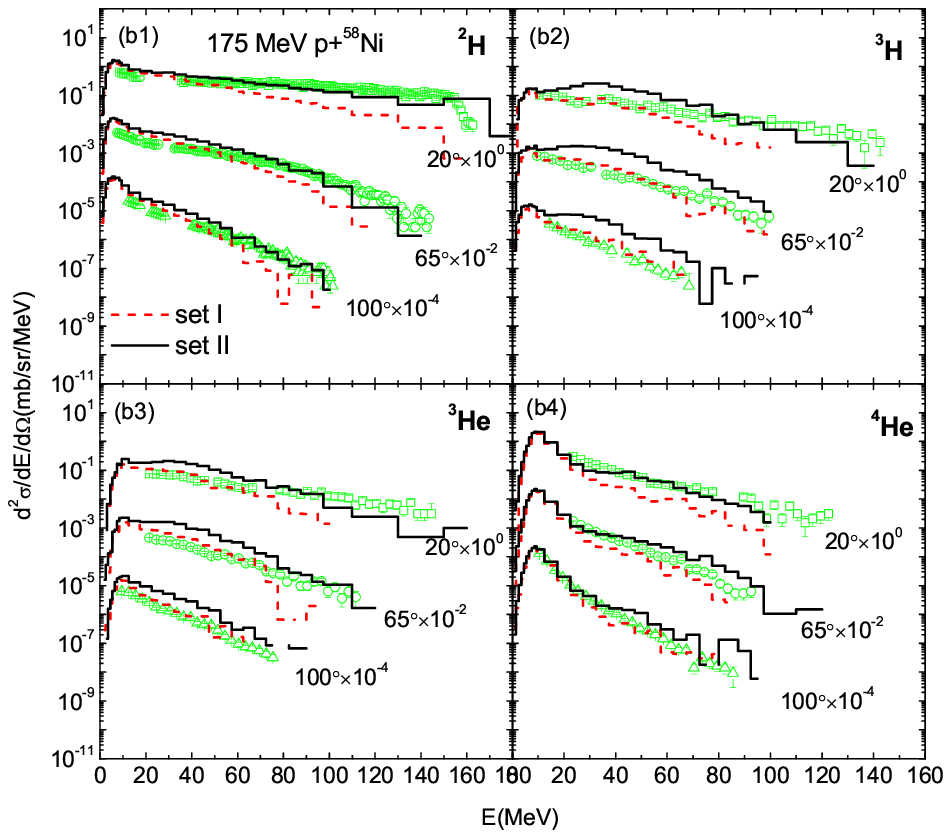}}
    \subfigure{\label{fig3c}
    \includegraphics[height=4.5cm]{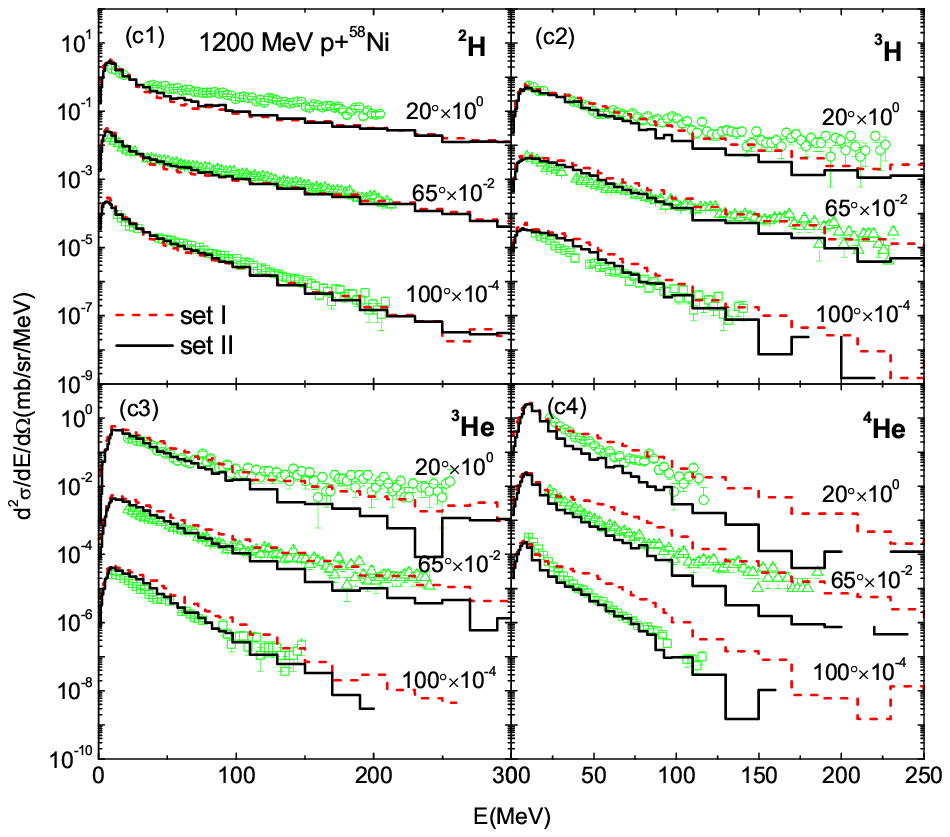}}
    \caption{(Color online) Calculated DDXs of LCPs
    in the reactions $n+^{56}$Fe at 175 MeV (experimental data are taken from \cite{Bevilacqwa}), and
    $p+^{58}$Ni at 175, 1200 MeV (experimental data are taken from \cite{Budzanowski,Budza2})
    with parameter setI and setII adopted, respectively.}
    \label{fig3}
    \end{center}
    \end{figure*}

\begin{figure*}[htp]
    \begin{center}
    \subfigure{\label{fig3a}
    \includegraphics[height=10cm]{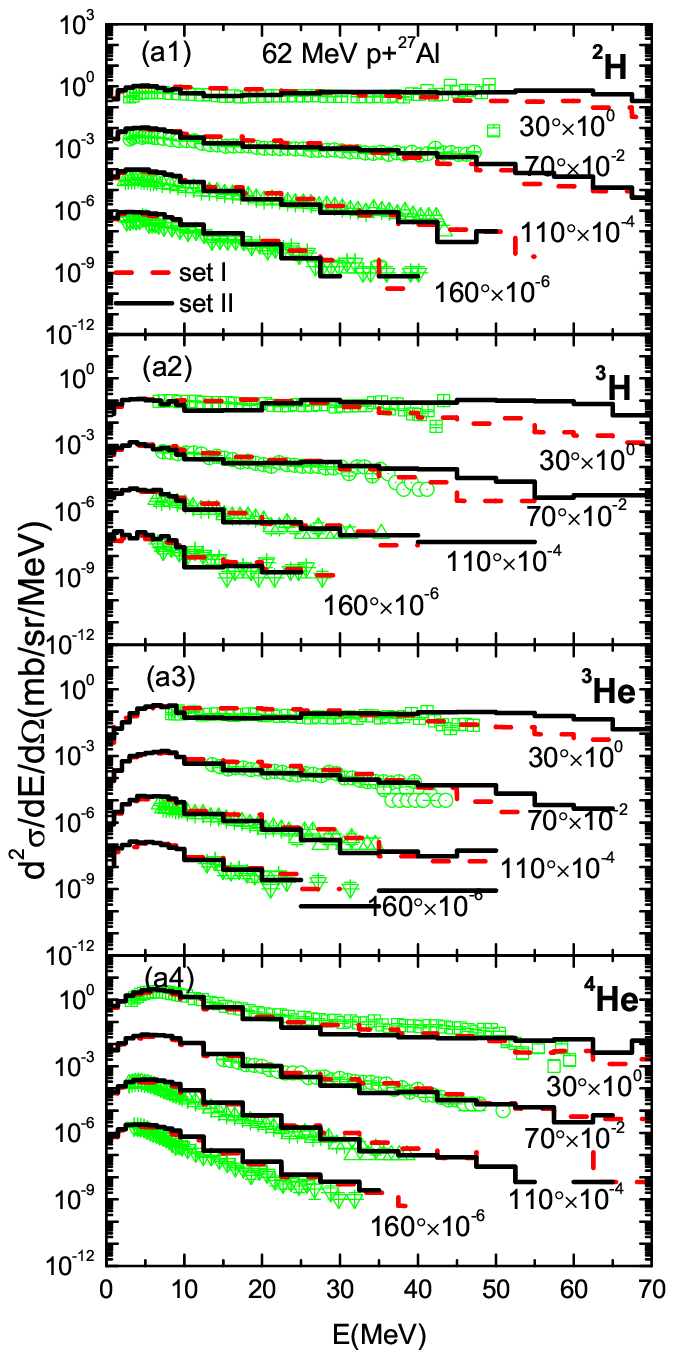}}
    \subfigure{\label{fig3b}
    \includegraphics[height=10cm]{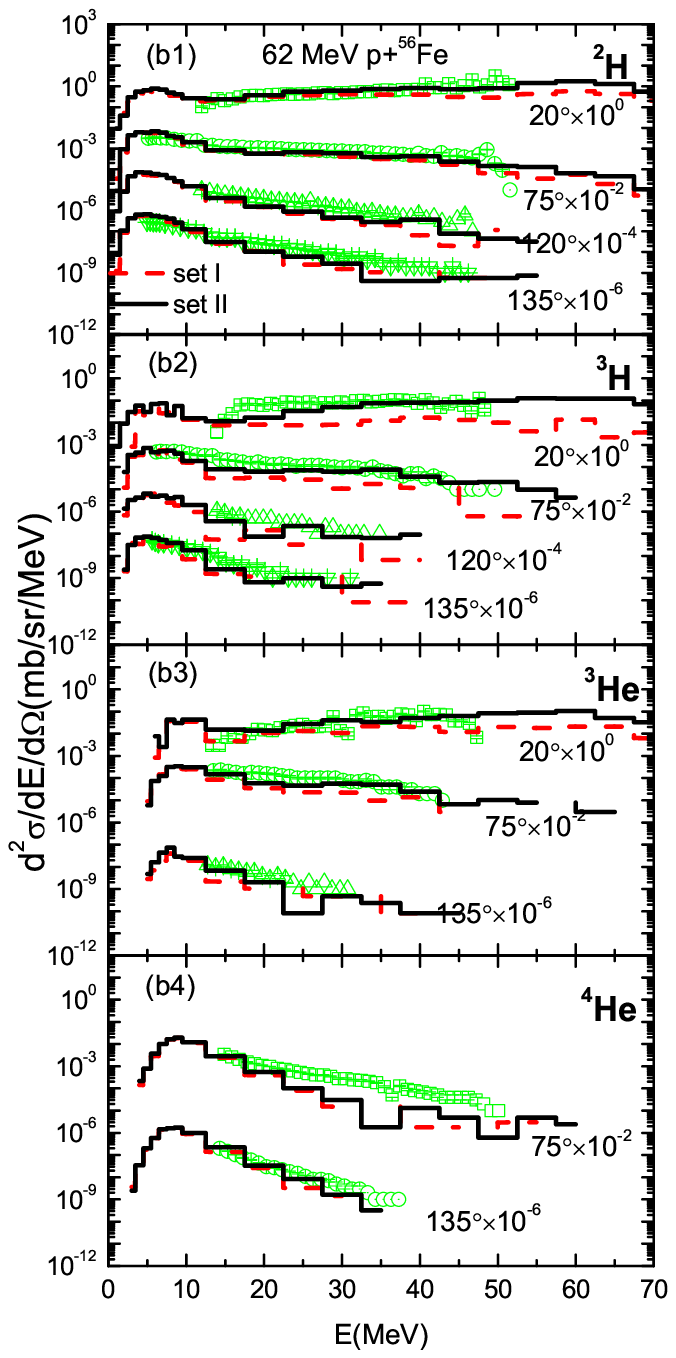}}
    \subfigure{\label{fig3c}
    \includegraphics[height=10cm]{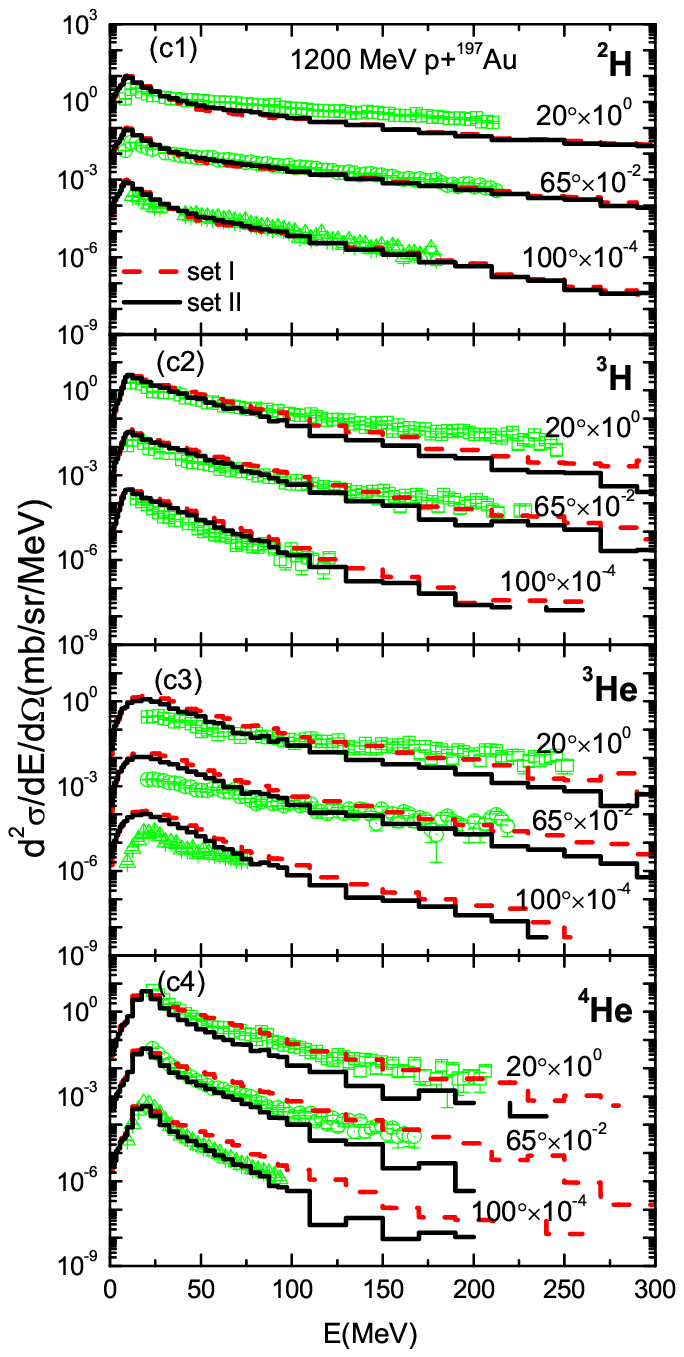}}
    \caption{(Color online) Calculated DDXs for LCPs
    in the reactions $p+^{27}$Al, $p+^{56}$Fe (middle pane) at 62 MeV,
    (experimental data are taken from \cite{Bert4}), and $p+^{197}$Au at 1200 MeV
    (experimental data are taken from \cite{Budza})
    with parameter setI and setII adopted, respectively.}
    \label{fig4}
    \end{center}
    \end{figure*}

    Figures \ref{fig2}-\ref{fig4} represent calculated DDXs of LCPs in several nucleon-reduced reactions
    with parameter setI and setII adopted, respectively.
    One can see that, although for some LCPs, for example, $^3$H in 175 MeV $p+^{58}$Ni
    and $^4$He in 1200 MeV $p+^{197}$Au reaction, the results calculated with parameter setII
    are not as good as the results calculated with parameter setI,
    the most experimental data can be described better with parameter setII.
    Because by using parameter setII, the independent phase space parameters for each LCP
    is helpful to rather satisfactorily fit the experimental data.

\begin{figure*}[htp]
    \begin{center}
    \includegraphics[height=10cm]{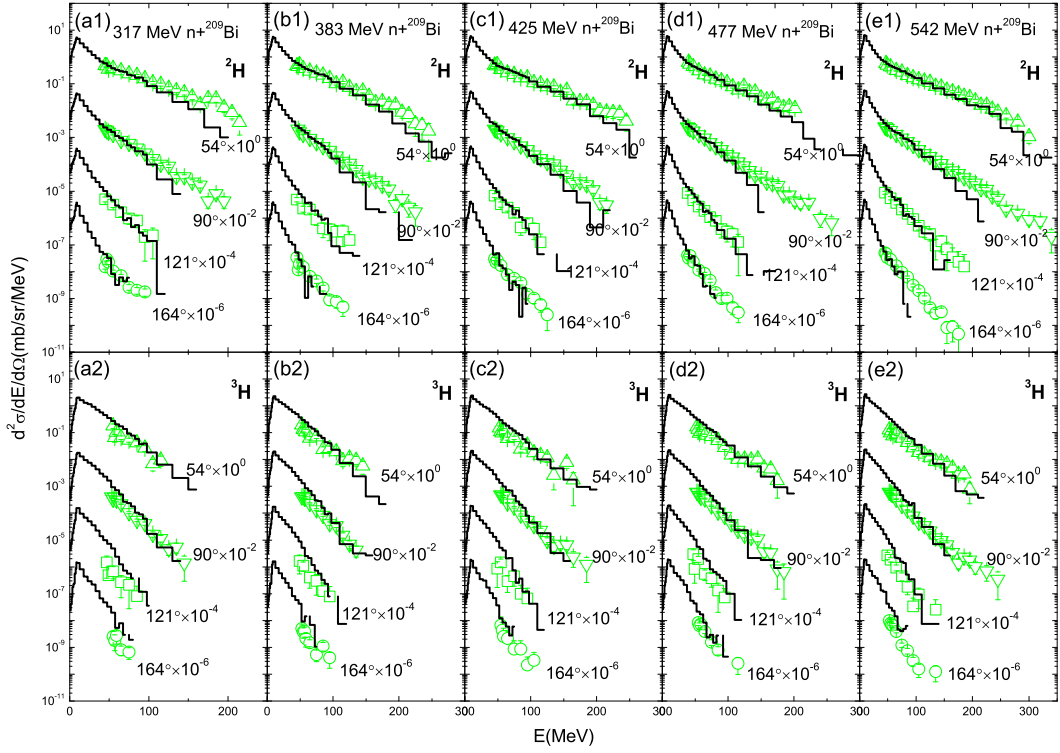}
    \caption{(Color online) Calculated DDXs of light complex particles
    produced in the reaction $n+^{209}$Bi and $n+^{209}$Bi at 317, 383, 425, 477 and 542 MeV, respectively.
    Experimental data are taken from \cite{Franz}.}
    \label{fig5}
    \end{center}
    \end{figure*}

    Finally, reactions of 317, 383, 425, 477, 542 MeV neutrons hit $^{209}$Bi are used to test the model.
    The comparisons between calculated DDXs of LCPs and experimental data are illustrated in Fig.~\ref{fig5}.
    From the figure, one can see that excellent agreement of LCPs DDXs are achieved.

\section{summary}

    In our last work, a phenomenological surface coalescence mechanism is introduced into ImQMD05 model.
    With surface coalescence mechanism introduced, the description on the light complex particles emission
    in nucleon-induced reactions is great improved,
    the experimental data of double differential cross sections of LCPs can be reproduced well.
    In this work, the phase space parameters in the surface coalescence model are readjusted
    with specific binding energy and separation energy effects for various LCPs being considered.
    The experimental data are better accounted with new parameters set.
    And what more important is that the new parameters set has better physical fundament.

    \begin{acknowledgments}
    This work was supported by the National Natural Science Foundation of China
    under Grant Nos.
    11005022, 
    11365004, 
    11365005, 
    11275052, 
    and by the Doctor Startup Foundation of Guangxi Normal University.

    \end{acknowledgments}



\end{document}